\documentclass[prl,aps,preprint,amsmath,docs,bm]{revtex4}

\begin{document}
\title{Gravitational Vacuum Condensate Stars\\ \ \\}
\author{Pawel O. Mazur}
\affiliation{Department of Physics and Astronomy, University of South
Carolina, Columbia, SC 29208, USA}
\author{Emil Mottola}
\affiliation{Theoretical Division, T-8, Los Alamos National
Laboratory,
MS B285, Los Alamos, NM 87545, USA\\ \ \\}

\begin{abstract}
{\small{A new final state of gravitational collapse is proposed.
By extending the concept of Bose-Einstein condensation to gravitational systems,
a cold, dark, compact object with an interior de Sitter condensate $p_{_V} = -\rho_{_V}$
and an exterior Schwarzschild geometry of arbitrary total mass $M$ is constructed. These are
separated by a shell with a small but finite proper thickness $\ell$ of fluid with eq. of
state $p=+\rho$, replacing both the Schwarzschild and de Sitter classical horizons.
The new solution has no singularities, no event horizons, and a global time. Its entropy
is maximized under small fluctuations and is given by the standard hydrodynamic entropy
of the thin shell, which is of order $k_{_B}\ell Mc/\hbar$, instead of the Bekenstein-Hawking
entropy formula, $S_{_{BH}}= 4\pi k_{_B} G M^2/\hbar c$. Hence unlike black holes, the new
solution is thermodynamically stable and has no information paradox.}}
\end{abstract}
\maketitle

{\bf Introduction.} Cold superdense stars with a mass above some critical value
undergo rapid gravitational collapse. Due to the impossibility
of halting this collapse by any known equation of state for high density matter,
a kind of consensus has developed that a collapsing star must inevitably arrive in a
finite proper time at a singular condition, called a black hole.

The characteristic feature of a black hole is its event horizon, the null surface of
finite area at which outwardly directed light rays hover indefinitely. For simplicity,
consider an uncharged, non-rotating Schwarzschild black hole with the static, spherically
symmetric line element,
\begin{equation}
ds^2 = -f(r)\, dt^2 + {dr^2 \over h(r)} + r^2\left( d\theta^2 + \sin^2\theta\,d\phi^2\right)
\,.
\label{sphsta}
\end{equation}
The functions $f(r)$ and $h(r)$ are equal in this case, and
\begin{equation}
f_{_S}(r) = h_{_S}(r) = 1 - {r_{_S}\over r}\,, \qquad r_{_S} \equiv \frac{2GM}{c^2}\,.
\label{Sch}
\end{equation}
At the event horizon, $r=r_{_S}$, the metric (\ref{sphsta}) becomes
singular. Since local curvature invariants remain regular at $r=r_{_S}$, a test particle
falling through the horizon experiences nothing catastrophic there (if $M$ is large
enough), and it is possible to find regular coordinates which analytically extend the
exterior Schwarzschild geometry through the event horizon into the interior region.

It is important to recognize that this mathematical procedure of analytic continuation through
a null hypersurface involves a physical assumption, namely that the stress-energy tensor
is vanishing there. Even in the classical theory, the hyperbolic character of the Einstein
equations allows generically for sources and discontinuities on the horizon which would violate
this assumption. Whether such analytic continuation is mandatory, or even permissable
in a more complete theory taking quantum effects into account, is still less certain.

Non-analytic behavior is typical of quantum many-body systems at a phase transition.
Quantum systems also exhibit macroscopic coherence effects which do not depend on local
forces becoming large. Thus the fact that the tidal forces on classical test bodies
falling through the event horizon are arbitrarily weak (proportional to
$M/r_{_S}^3 \sim 1/M^2$) for an arbitrarily large black hole does not imply that quantum
effects must be unimportant there. Electron waves restricted to the region
outside an Aharonov-Bohm solenoid, where the electromagnetic field strength vanishes and
there are no classical forces whatsoever, nevertheless experience a shift in their
interference fringes. Qualitatively new effects such as these arise because quantum matter
has extended wavelike properties, with many-body statistical correlations that can in no way
be captured by consideration of pointlike test particles responding only to local forces.

A photon with asymptotic frequency $\omega$ and energy $\hbar \omega$
far from the black hole has a local energy $\hbar\omega\,f^{-{1\over 2}}$,
which diverges at the event horizon. Unlike classical test particles, when
$\hbar \neq 0$ such extremely blue shifted photons are necessarily present in the vacuum
as virtual quanta. Their effects upon the geometry depend upon the quantum state of the vacuum,
defined by boundary conditions on the wave equation in a non-local way over all of space,
and $\langle T_a^{\ b} \rangle$ may be large at $r=r_{_S}$, notwithstanding
the smallness of the local classical curvature there \cite{Bou75,Bou76,CFul}. Since the limits
$r\rightarrow r_{_S}$ and $\hbar \rightarrow 0$ do not commute, non-analytic behavior near
the event horizon, quite different from that in the strictly classical ($\hbar \equiv 0$)
situation is possible in the quantum theory.

The non-analytic nature of the classical limit $\hbar \rightarrow 0$ may be seen also
in the thermodynamic analogy for the laws of black hole mechanics. Pursuing the
analogy of these classical laws with thermodynamics, Bekenstein suggested that black hole event
horizons carry an intrinsic entropy proportional to their area, $4\pi r_{_S}^2$ \cite{Bek72,Bek73,Bek74}.
In order to have the units of entropy, the horizon area must be multiplied by a constant with units
of $c^3 k_{_B}/\hbar G$. When Hawking found that a flux of radiation could be emitted from a
black hole with a well-defined temperature, $T_{_H} = \hbar c^3/8\pi k_{_B} GM$ \cite{Haw74,Haw75},
the thermodynamic analogy of area with entropy received some support, since the conservation of
energy in the system can be written in the form,
\begin{equation}
dE = T_{_H} d S_{_{BH}} = \frac{\hbar c^3}{8\pi G k_{_B} M}\
d\left(\frac{4\pi k_{_B}GM^2}{\hbar c}\right)\,,
\label{sbh}
\end{equation}
suggestive of the first law of thermodynamics.

The curious feature of relation (\ref{sbh}) is that $\hbar$ cancels out, and plays no
dynamical role. The identification of $S_{_{BH}}$ with the entropy of the black hole is
founded on the dynamics of purely classical relativity ({\it i.e.} $\hbar = 0$), through
the area law of refs. \cite{Chr,ChrRuf,HawEll}. If this identification of rescaled classical
area with entropy is to be valid in the quantum theory, then the limit $\hbar \rightarrow 0$
(with $M$ fixed) which yields an arbitrarily low temperature, would assign to the black hole
an arbitrarily large entropy, completely unlike the zero temperature limit of any other cold
quantum system.

Closely related to this paradoxical result is the fact, also pointed out by Hawking \cite{Haw76},
that a temperature inversely proportional to $M= E/c^2$ implies a negative heat capacity:
$\frac{dE}{dT} = -(\hbar c^5/8\pi G k_{B}) \,T^{-2}<0$. However, a negative heat capacity is
impossible for any system in stable equilibrium, since the heat capacity is proportional
to the square of the energy fluctuations of the system. If this quantity is negative,
the system cannot be in stable equilibrium at all, and the applicability of
equilibrium thermodynamic relations is questionable. Attempts to evaluate the entropy of an
uncharged black hole directly from statistical considerations ($S= k_{_B} \ln \Omega$)
produce divergent results due to the unbounded number of wave modes infinitely close to
the event horizon \cite{tH85,BKLS}. The entropy of fields in a fixed Schwarzschild background
would also be expected to scale linearly with the number $N$ of independent fields,
whereas $S_{_{BH}}$ is independent of $N$.

Ignoring these myriad difficulties and nevertheless interpreting (\ref{sbh}) literally
as a thermodynamic relation implies that a black hole has an enormous entropy,
$S_{_{BH}}\simeq 10^{77} k_{_B}(M/M_{\odot})^2$, far in excess of a typical
stellar progenitor with a comparable mass. The associated `information paradox' and implied
violation of unitarity has been characterized as so serious as to require an alteration
in the principles of quantum mechanics \cite{Haw82,tH99}.

This paradoxical state of affairs arising from an area law originally derived in a strictly
classical framework, together with the cancellation of $\hbar$ in (\ref{sbh}), suggest that
(\ref{sbh}) may not be a generally valid quantum relation at all, but only the (improper)
classical limit of such a relation. The cancellation of $\hbar$ here is reminiscient of a
similar cancellation in the energy density of modes of the radiation field in thermodynamic
equilibrium, $\hbar \omega\, n (\omega)\, \omega^2 d\omega \rightarrow k_{_B}T \omega^2 d\omega$
in the Rayleigh-Jeans limit of very low frequencies, $\hbar \omega \ll k_{_B}T$. Improper extension
of this low energy relation from Maxwell theory into the quantum high frequency regime leads to an
ultraviolet catastrophe, similar to that encountered in the counting of wave modes near an event
horizon. Conversely, treating the atoms in a solid as classical point particles leads to
equipartition of energy and an incorrect prediction (the Dulong-Petit law) of constant specific
heat for crystals at low temperatures. Both of these difficulties precipitated and were resolved
by the rise of a new quantum theory of matter and radiation.

In the illustrative case of the Einstein single frequency crystal, the difference between
the heat capacity $C_{_V}(T)$ at the temperature $T$  and its Dulong-Petit high temperature
limit, $C_{_V}(\infty)$ is $\Delta C_{_V}(T) = C_{_V}(T) - C_{_V}(\infty)$, which has the
leading high temperature behavior $\Delta C_{_V}(T) \propto - T^{-2}$, exactly the same
temperature dependence as the negative Bekenstein-Hawking heat capacity of the Schwarzschild
black hole. When the term $C_{_V}(\infty)$ arising from the discrete atomistic degrees of freedom
is restored, the true heat capacity $C_{_V}(T)$ of the crystal is positive. Application of the
Einstein energy fluctuation formula to a black hole shows that the Bekenstein-Hawking heat capacity
is identical to what one would obtain for a large number of weakly interacting massive bosons, each
with a mass of order of the Planck mass, close to the Dulong-Petit limit, but where the term
analogous to $C_{_V}(\infty)$ has been dropped \cite{Maz}.

A related consideration suggesting that the Hawking emission depends on an improper
ultraviolet extrapolation of classical waves into the quantum regime comes
from examining the origin of these waves in the past. Although the Hawking temperature
$T_{_H}$ is very small for a large black hole, simple kinematic considerations show
that the Hawking modes observed at a time $t$ long after the formation of the hole originate as
incoming vacuum modes at a greatly blue shifted frequency,
$\omega_{in} \sim (k_{_B}T_{_H}/\hbar)\,\exp(ct/2r_{_S})$.
In other words, the calculation of the Hawking flux at late times assumes that the local
properties of the fixed spacetime geometry are known with arbitrarily high precision, even at
exponentially sub-Planck length and time scales, where one would normally question the
semi-classical approximation, and indeed the existence of any well-defined metric at all. The
contribution to the local stress-energy tensor, $\langle T_a^{\ b} \rangle$ of these highly
blue shifted modes diverges at the horizon, which is another form of the ultraviolet catastrophe.
Only an exact balance with time-reversed highly blue shifted ingoing modes can cancel this
divergence in the precisely tuned Hartle-Hawking thermal state \cite{HarHaw}. This state is in
any case unstable due to its negative heat capacity. In any other thermal state, with
$T\neq T_{_H}$, such as would be expected to arise from fluctuations, the divergence at the horizon
is not cancelled, and one must expect a substantial quantum backreaction on the geometry there,
which invalidates the basic assumption of an arbitrarily accurately known fixed classical
Schwarzschild background, arbitrarily close to $r= r_{_S}$.

Note that since it transforms as a tensor under coordinate transformations, a large local
$\langle T_a^{\ b} \rangle$ near $r= r_{_S}$ is perfectly consistent with the Equivalence
Principle, except in its strongest possible form which would admit no non-local effects of
any kind, including those of macroscopic coherence and entanglement, known to exist in both
relativistic and non-relativistic quantum many-body systems.

In earlier models of backreaction the black hole was immersed in a Hawking radiation atmosphere,
with an effective equation of state, $p =\kappa \rho$. It was found that due to the blue shift
effect, the backreaction of such an atmosphere on the metric near $r=r_{_S}$ is enormous, with the
interior region quite different from the vacuum Schwarzschild solution, and with a large entropy
of order of $S_{_{BH}}$ from the fluid alone \cite{ZP}. In fact,
$S = 4\,{\kappa + 1\over 7 \kappa + 1} S_{_{BH}}$, becoming equal to the Bekenstein-Hawking
entropy for $\kappa = 1$. Aside for accounting for the entropy $S_{_{BH}}$ from purely standard
hydrodynamical relations, this result suggests that the maximally stiff equation of state
consistent with the causal limit may play a role in the quantum theory of fully collapsed
objects. Despite such very suggestive features indicating the importance of backreaction
on the geometry, these models cannot be viewed as a satisfactory solution to the final state
of the collapse problem, since they involve huge (Planckian) energy densities near $r_{_S}$,
as well as a negative mass singularity at $r=0$. The negative mass singularity
arises because a repulsive core is necessary to counteract the self-attractive gravitation
of the dense relativistic fluid with positive energy.

Recently another proposal for incorporating quantum effects near the
horizon has been made \cite{CHLS01,Lau,CHLS03}, with a critical surface of a quantum phase transition
replacing the classical horizon, and the interior replaced by a region with eq. of state,
$p = - \rho < 0$. Equivalent to a positive cosmological term in Einstein's eqs., this eq. of
state was first proposed for the final state of gravitational collapse
by Gliner \cite{Gli} and considered in a cosmological context by Sakharov \cite{Sak}.
It violates the strong energy condition $\rho + 3p \ge 0$ used in proving the classical
singularity theorems. As is now well known because of the observations of distant supernovae
implying an accelerating universe, the geodesic worldlines of test particles in a region of
spacetime where $\rho + 3p < 0$ diverge from each other, mimicking the effects of a
repulsive gravitational potential. Thus such a region free of any singularities in
the interior region $r < r_{_S}$ can replace the unphysical negative mass singularity
encountered in the pure $\kappa =1$ fluid model.

Based upon these considerations, in this paper we show that a consistent static solution
of Einstein's equations can be constructed, with the critical surface of refs.
\cite{CHLS01,CHLS03} replaced by a thin shell of ultra-relativistic fluid with eq. of state
$p=\rho $. Because of its vacuum energy interior with $p=-\rho$ the new solution is
stable and free of all singularities. Its entropy is a local maximum of the hydrodynamic entropy,
whose modest value given by (\ref{entsh})-(\ref{entest}) below is easily attainable in a physical
collapse process from a stellar progenitor with a comparable mass.

The model we arrive at is that of a low temperature condensate of weakly interacting massive
bosons trapped in a self-consistently generated cavity, whose boundary layer can be described
by a thin shell. The assumption required for a solution of
this kind to exist is that gravity, {\it i.e.} spacetime itself, must undergo a quantum
vacuum rearrangement phase transition in the vicinity of $r=r_{_S}$. In
this region quantum zero-point fluctuations dominate the stress-energy tensor,
and become large enough to influence the geometry, regardless of the
composition of the matter undergoing the gravitational collapse. As the causal limit
$p=\rho$ is reached, the interior spacetime becomes unstable to the formation of
a gravitational Bose-Einstein condensate (GBEC), described by a non-zero macroscopic
order parameter in the effective low energy description. Since a condensate is a single macroscopic
quantum state with zero entropy, a model of a cold condensate repulsive core as the stable,
non-singular endpoint of gravitational collapse provides a resolution of the information paradox,
which is completely consistent with quantum principles \cite{Chap}. The interior and exterior
regions are separated by a thin surface layer near $r = r_{_S}$ where the vacuum condensate
disorders. Any entropy in the configuration can reside only in the excitations of this boundary
layer. A suggestion for the effective theory incorporating the effects of quantum anomalies
that describes this fully disordered phase, where the role of the order parameter is played
by the conformal part of the metric has been presented elsewhere \cite{MMPRD}. For a recent
review of investigations of other non-singular quasi-black-hole (QBH) models see \cite{Dym}.

{\bf The Vacuum Condensate Model.} In an effective mean field treatment for a perfect fluid at rest
in the coordinates (\ref{sphsta}), any static, spherically symmetric collapsed object
must satisfy the Einstein eqs. (in units where $c=1$),
\begin{equation}
-G_{\ t}^t = {1\over r^2} {d\over dr}\left[r\left(1 - h\right)\right]
= -8\pi G\, T_{\ t}^t = 8\pi G \rho\,,
\label{Einsa}
\end{equation}
\begin{equation}
G_{\ r}^r = {h\over r f} {d f\over dr}  + {1\over r^2} \left(h -1\right) =
8\pi G \,T_{\ r}^r = 8\pi G p\,,
\label{Einsb}
\end{equation}
\vspace{-.2cm}
\noindent
together with the conservation eq.,
\begin{equation}
\nabla_a T^a_{\ r} = {d p\over dr} + {\rho + p\over 2f} \,{d f\over dr}
+ 2\, \frac{ p-p_{\perp}}{r}= 0\,,
\label{cons}
\end{equation}
which ensures that the other components of the Einstein eqs. are satisfied.
In the general spherically symmetric situation the tangential pressure,
$p_{\perp} \equiv T^{\theta}_{\ \theta} = T^{\phi}_{\ \phi}$ is not necessarily
equal to the radial normal pressure $p = T^r_{\ r}$. However, for purposes of
developing the simplest possibility first, we restrict ourselves in this paper
to the isotropic case where $p_{\perp} = p$. In that case, we have three first
order eqs. for four unknown functions of $r$, {\it viz.} $f, h, \rho$, and $p$.
The system becomes closed when an eq. of state for the fluid, relating $p$ and $\rho$
is specified. Because of the considerations of the Introduction we allow for three different
regions with the three different eqs. of state,
\begin{equation}
\begin{array}{clcl}
{\rm I.}\ & {\rm Interior:}\ & 0 \le r < r_1\,,\ &\rho = - p \,,\\
{\rm II.}\ & {\rm Thin\ Shell:}\ & r_1 < r < r_2\,,\ &\rho = + p\,,\\
{\rm III.}\ & {\rm Exterior:}\ & r_2 < r\,,\ &\rho = p = 0\,.
\end{array}
\end{equation}

In the interior region $\rho = - p$ is a constant from
(\ref{cons}). Let us call this constant $\rho_{_V} = 3H_0^2/8\pi G$. If we
require that the origin is free of any mass singularity then the
interior is determined to be a region of de Sitter spacetime in static coordinates, {\it i.e.}
\begin{equation}
{\rm I.}\qquad f(r) = C\,h(r) = C\,(1 - H_0^2\,r^2)\,,\quad 0 \le r \le r_1\,,
\label{deS}
\end{equation}
where $C$ is an arbitrary constant, corresponding to the freedom to redefine the
interior time coordinate.

The unique solution in the exterior vacuum region which approaches flat spacetime
as $r \rightarrow \infty$ is a region of Schwarzschild spacetime (\ref{Sch}),
{\it viz.}
\begin{equation}
{\rm III.} \qquad f(r) = h(r) = 1 - {2GM\over r}\,,\qquad  r_2 \le r\,.
\end{equation}
The integration constant $M$ is the total mass of the object.

The only non-vacuum region is region II. Let us define the dimensionless
variable $w$ by $w\equiv 8\pi G r^2 p$, so that
eqs. (\ref{Einsa})-(\ref{cons}) with $\rho = p$ may be recast in the form,

\begin{equation}
{dr\over r} = \ {dh \over 1-w-h}\,,
\label{ueqa}
\end{equation}
\begin{equation}
{dh\over h} = -{1-w-h \over 1 + w - 3h}\, {dw\over w}\,,
\label{ueqb}
\end{equation}
\vspace{-.2cm}

\noindent
together with $p f \propto wf/r^2$ a constant.
Eq. (\ref{ueqa}) is equivalent to the definition of the (rescaled)
Tolman mass function by $h = 1 - 2m(r)/r$ and $d m(r) = 4\pi G\, \rho r^2\, dr = w\, dr/2$
within the shell. Eq. (\ref{ueqb}) can be solved only numerically in general. However,
it is possible to obtain an analytic solution in the thin shell limit, $0 < h \ll 1$, for
in this limit we can set $h$ to zero on the right side of (\ref{ueqb}) to leading order,
and integrate it immediately to obtain
\begin{equation}
h \equiv 1- {2m\over r} \simeq  \epsilon\ {(1 + w)^2\over w \ }\ll 1\,,
\label{hshell}
\end{equation}
in region II, where $\epsilon$ is an integration constant.
Because of the condition $h \ll 1$, we require $\epsilon \ll 1$, with $w$
of order unity. Making use of eqs. (\ref{ueqa})-(\ref{ueqb}) and (\ref{hshell}) we have
\begin{equation}
{dr \over r} \simeq - \epsilon\, dw\, {(1 + w)\over w^2}\,.
\label{req}
\end{equation}
Because of the approximation $\epsilon \ll 1$, the radius $r$ hardly changes within
region II, and $dr$ is of order $\epsilon \,dw$. The final unknown function $f$ is given
by $f = (r/r_1)^2 (w_1/w) f(r_1) \simeq  (w_1/w) f(r_1)$ for small $\epsilon$, showing that
$f$ is also of order $\epsilon$ everywhere within region II and its boundaries.

At each of the two interfaces at $r=r_1$ and $r=r_2$ the induced three dimensional
metric must be continuous. Hence $r$ and $f(r)$ are continuous at the interfaces,
and
\begin{equation}
f(r_2) \simeq \frac{w_1}{w_2} f(r_1) = \frac{C w_1}{w_2} (1-H_0^2 r_1^2) = 1 - \frac{2GM}{r_2}\,.
\end{equation}
To leading order in $\epsilon \ll 1$ this relation implies that
\begin{equation}
r_1 \simeq \frac{1}{H_0} \simeq 2GM \simeq r_2\,.
\label{HMiden}
\end{equation}
Thus the interfaces describing the phase boundaries at $r_1$ and $r_2$ are very close to
the classical event horizons of the de Sitter interior and the Schwarzschild exterior.

The significance of $0< \epsilon \ll 1$ is that both $f$ and $h$ are of order
$\epsilon$ in region II, but are nowhere vanishing. Hence there is no event horizon,
and $t$ is a global time. A photon experiences a very large, ${\cal O}(\epsilon^{-{1\over 2}})$
but finite blue shift in falling into the shell from infinity.
The proper thickness of the shell between these interface boundaries is
\begin{equation}
\ell = \int_{r_1}^{r_2}\, dr\,h^{-{1\over 2}}  \simeq r_{_S}
\epsilon^{1\over 2}\int_{w_2}^{w_1} dw\, w^{-{3\over 2}}
\sim \epsilon^{1\over 2} r_{_S}\,,
\label{thickness}
\end{equation}
and very small for $\epsilon \rightarrow 0$. Because of (\ref{HMiden}) the
constant vacuum energy density in the interior is just the total mass $M$
divided by the volume, {\it i.e.} $\rho_{_V} \simeq 3M/4\pi r_{_S}^3$,
to leading order in $\epsilon$. The energy within the shell itself,
\begin{equation}
E_{\rm II} = 4\pi \int_{r_1}^{r_2}\rho\,r^2 dr \simeq \epsilon M
\int_{w_2}^{w_1}{dw\over w}\,(1 + w)\sim \epsilon M
\end{equation}
is extremely small.

We can estimate the size of $\epsilon$ and $\ell$ by consideration
of the expectation value of the quantum stress tensor in the static exterior
Schwarzschild spacetime. In the static vacuum state corresponding to no incoming
or outgoing quanta at large distances from the object, {\it i.e.} the Boulware
vacuum \cite{Bou75,Bou76}, the stress tensor near $r = r_{_S}$ is the negative of the stress tensor of
massless radiation at the blue shifted temperature, $T_{loc} = T_{_H}/\sqrt {f(r)}$
and diverges as $T_{loc}^4 \sim f^{-2}(r)$ as $r\rightarrow r_{_S}$. The location
of the outer interface occurs at an $r$ where this local stress-energy $\propto M^{-4}
\epsilon^{-2}$, becomes large enough to affect the classical Schwarzschild curvature
$\sim M^{-2}$, {\it i.e.} when
\begin{equation}
\epsilon \sim \frac{M_{pl}} {M} \simeq 10^{-38}\ \left(\frac {M_{\odot}}{M}\right)\,,
\label{epsest}
\end{equation}
where $M_{Pl}$ is the Planck mass $\sqrt{\hbar c/G} \simeq 2 \times 10^{-5}$ gm.
Thus $\epsilon$ is indeed very small for a stellar mass object, justifying the
approximation {\it a posteriori}. With this semi-classical estimate for $\epsilon$
we find
\begin{equation}
\ell \sim \,\sqrt {L_{_{Pl}}\, r_{_S}} \simeq \, 3 \times 10^{-14}\,
\left(\frac {M} {M_{\odot}}\right)^{1\over 2} {\rm cm.}
\label{meanell}
\end{equation}
Although still microscopic, the thickness of the shell is very much larger than the Planck
scale $L_{Pl} \simeq 2 \times 10^{-33}$ cm. The energy density and pressure in the shell
are of order $M^{-2}$ and far below Planckian for $M\gg M_{Pl}$, so that the geometry can be
described reliably by Einstein's equations in both regions I and II.

Although $f(r)$ is continuous across the interfaces at $r_1$ and $r_2$, the discontinuity
in the eqs. of state does lead to discontinuities in $h(r)$ and the first derivative of $f(r)$
in general. Defining the outwardly directed unit normal vector to the interfaces,
$n^b= \delta_r^{\ b} \sqrt{h(r)}$, and the extrinsic curvature $K_a^{\ b} = \nabla_an^b$,
the Israel junction conditions determine the surface stress energy $\eta$ and surface tension
$\sigma$ on the interfaces to be given by the discontinuities in the extrinsic curvature through
\cite{Isr}
\begin{equation}
[K_t^{\ t}] = \left[ \frac {\sqrt h}{2f} \frac {df}{dr} \right] = 4\pi G(\eta - 2\sigma)\,,
\label{junca}
\end{equation}
\begin{equation}
\left[K_{\theta}^{\ \theta}\right] = [K_{\phi}^{\ \phi}] = \left[ \frac {\sqrt h}{r}\right]
= - 4\pi G\eta\,.
\label{juncb}
\end{equation}
\noindent
Since $h$ and its discontinuities are of order $\epsilon$, the energy density in the surfaces
$\eta \sim \epsilon^{\frac{1}{2}}$, while the surface tensions are of order $\epsilon^{-\frac{1}{2}}$.
The simplest possibility for matching the regions is to require that the surface energy densities
on each interface vanish. From (\ref{juncb}) this condition implies that $h(r)$
is also continuous across the interfaces, which yields the relations,
\begin{eqnarray}
h(r_1) &=& 1-H_0^2 r_1^2 \simeq \epsilon \,{(1+ w_1)^2 \over
w_1}\,,
\label{matcha}\\
h(r_2) &=& 1 - {2GM\over r_2} \simeq  \epsilon \,{(1+ w_2)^2 \over w_2}\,,\\
{f(r_2)\over h(r_2)} &=& 1 \simeq {w_1\over w_2} {f(r_1)\over h(r_2)} = C
\left({1+w_1\over 1+w_2}\right)^2
\label{matchb}
\end{eqnarray}
\noindent
From (\ref{req}) $dw/dr <0$, so that $w_2 < w_1$ and $C < 1$. In this case of vanishing
surface energies $\eta = 0$ the surface tensions are determined by (\ref{junca})-(\ref{juncb}) to be
\begin{eqnarray}
&&\quad \sigma_1 \simeq -{1\over 32\pi G^2M}
{(3 + w_1)\over (1 + w_1)}\left({w_1\over \epsilon}\right)^{1\over 2}\,,
\label{surfa}\\
&&\quad \sigma_2 \simeq {1\over
32\pi G^2M}  {w_2\over (1 + w_2)}\left({w_2\over \epsilon}\right)^{1\over 2}.
\label{surfb}
\end{eqnarray}
\noindent
to leading order in $\epsilon$ at $r_1$ and $r_2$ respectively. The negative surface tension
at the inner interface is equivalent to a positive tangential pressure, which implies
an outwardly directed force on the thin shell from the repulsive vacuum within. The positive
surface tension on the outer interfacial boundary coresponds to the more familiar case of an
inwardly directed force exerted on the thin shell from without.

The entropy of the configuration may be obtained from the Gibbs relation,
$p + \rho = sT + n\mu$, if the chemical potential $\mu$ is known in each region.
In the interior region I, $p + \rho = 0$ and the excitations are the usual
transverse gravitational waves of the Einstein theory in de Sitter space.
Hence the chemical potential $\mu$ may be taken to vanish and
the interior has zero entropy density $s=0$, consistent with a single
macroscopic condensate state, $S= k_{_B} \ln\Omega = 0$ for $\Omega =1$.
In region II there are several alternatives depending upon the nature
of the fundamental excitations there. The $p=\rho$ eq. of state may come
from thermal excitations with negligible $\mu$ or it may come from a
conserved number density $n$ of gravitational quanta at zero temperature.
Let us consider the limiting case of vanishing $\mu$ first.

If the chemical potential can be neglected in region II, then the entropy of the shell
is obtained from the eq. of state, $p = \rho = (a^2/8\pi G) (k_{_B} T/\hbar)^2$. The
$T^2$ temperature dependence follows from the Gibbs relation with $\mu =0$, together
with the local form of the first law $d\rho = T ds$. The Newtonian constant $G$ has
been introduced for dimensional reasons and $a$ is a dimensionless constant.
Using the Gibbs relation again the local specific entropy density
$s(r) = a^2k_{_B}^2 T(r)/4\pi\hbar^2 G = a(k_{_B}/\hbar)(p/2\pi G)^{1\over 2}$
for local temperature $T(r)$. Converting to our previous variable $w$, we find
$s = (ak_{_B}/4\pi\hbar Gr)\,w^{1\over 2}$ and the entropy of the fluid within the shell is
\begin{equation}
S =4\pi \int_{r_1}^{r_2}s\,r^2\,dr\,h^{-{1\over 2}}\simeq
{ak_{_B}r_{_S}^2 \over \hbar G}\
\epsilon^{1\over 2} \ \ln \left({w_1\over w_2}\right)\,,
\label{entsh}
\end{equation}
to leading order in $\epsilon$. Using (\ref{thickness}) and (\ref{meanell}), this is
\begin{equation}
S \sim a\, k_{_B}{M\ell \over \hbar}
\sim 10^{57}\ a\,k{_{_B}}\,\left(\frac {M} {M_{\odot}}\right)^{3\over 2}\ll S_{_{BH}}\,.
\label{entest}
\end{equation}
The maximum entropy of the shell and therefore of the entire configuration is some $20$
orders of magnitude smaller than the Bekenstein-Hawking entropy for a solar mass object,
and of the same order of magnitude as a typical progenitor of a few solar masses.
The scaling of (\ref{entest}) with $M^{\frac{3}{2}}$ is also the same as that for supermassive
stars with $M > 100\,M_{\odot}$, whose pressure is dominated by radiation pressure \cite{ZelNov}.
Thus the formation of the GBEC star from either a solar mass or supermassive stellar progenitor does
not require an enormous generation or removal of entropy, and there is no information paradox.

Because of the absence of an event horizon, the GBEC star does not emit Hawking radiation.
Since $w$ is of order unity in the shell while $r \simeq r_{_S}$, the {\it local} temperature
of the fluid within the shell is of order $T_{_H} \sim \hbar/k_{_B}GM$. The strongly
redshifted temperature observed at infinity is of order $\sqrt\epsilon\, T_{_H}$,
which is very small indeed. Hence the rate of any thermal emission from the shell is
negligible.

If we do allow for a positive chemical potential within the shell, $\mu >0$, then
the temperature and entropy estimates just given become upper bounds, and it is possible to
approach a zero temperature ground state with zero entropy. This non-singular final state
of gravitational collapse is a cold, completely dark object sustained against
any further collapse solely by quantum zero-point pressure.

{\bf Stability.} In order to be a physically realizable endpoint of gravitational
collapse, any quasi-black hole candidate must be stable \cite{Maz}.
Since only the region II is non-vacuum, with a `normal' fluid and a positive
heat capacity, it is clear that the solution is thermodynamically stable.
The most direct way to demonstrate this stability is to work in the microcanonical ensemble
(in the case of zero chemical potential) with fixed total $M$, and show that the entropy functional,
\begin{equation}
S= {a k_{_B}\over \hbar G\,}\int_{r_1}^{r_2}r \,dr\, \left(2\,{dm\over dr}\right)^{1\over 2}
\left(1 - {2m(r) \over r}\right)^{-{1\over 2}}\,,
\label{entropy}
\end{equation}
is maximized under all variations of $m (r)$ in region II with the endpoints $(r_1, r_2)$,
or equivalently $(w_1, w_2)$ fixed.

The first variation of this functional with the endpoints
$r_1$ and $r_2$ fixed vanishes, {\it i.e.} $\delta S =0$ by the Einstein eqs.
(\ref{Einsa})-(\ref{Einsb}) for a static, spherically symmetric star. Thus any solution of
eqs. (\ref{Einsa})-({\ref{cons}) is guaranteed to be an extremum of $S$ \cite{Coc}.
This is also consistent with regarding Einstein's eqs. as a form of hydrodynamics,
strictly valid only for the long wavelength, gapless excitations in gravity.
In the context of a hydrodynamic treatment, thermodynamic stability is also a
necessary and sufficient condition for the {\it dynamical} stability of a static,
spherically symmetric solution of Einstein's equations \cite{Coc}.

The second variation of (\ref{entropy}) is
\begin{equation}
\delta^2 S = {a k_{_B}\over \hbar G\,}\int_{r_1}^{r_2}r\,
dr\, \left(2{dm\over dr}\right)^{-{3\over 2}}
h^{-\frac{1}{2}}\,\left\{-\left[{d(\delta m)\over dr}\right]^2 + 2
{(\delta m)^2\over r^2h^2}{d m\over dr}\left(1+ 2{dm\over dr}\right)\right\}\,,
\label{varent}
\end{equation}
when evaluated on the solution.
Associated with this quadratic form in $\delta m$ is a second order linear differential
operator $\cal L$ of the Sturm-Liouville type, {\it viz.}
\begin{equation}
{\cal L} \chi \equiv {d\over dr} \left\{r h^{-{1\over 2}} \left({d m\over dr}\right)^{-{3\over 2}}
{d\chi\over dr}\right\} +\,{2\over r}\,h^{-{5\over 2}}
\left({dm\over dr}\right)^{-{1\over 2}}\left(1+ 2\,{dm\over dr}\right)\chi \,.
\label{stli}
\end{equation}
This operator possesses two solutions satisfying ${\cal L}\chi_{_0} = 0$, obtained by
variation of the classical solution, $m (r; r_1, r_2)$ with respect to the parameters
$(r_1, r_2)$. Since these correspond to varying the positions of the interfaces, $\chi_{_0}$
does not vanish at $(r_1, r_2)$ and neither function is a true zero mode. For example, it is easily
verified that one solution is $\chi_{_0} = 1 - w$, from which the second linearly independent solution
$(1-w)\ln\,w + 4$ may be obtained. For any linear combination of these we may set $\delta m \equiv
\chi_{_0}\, \psi$, where $\psi$ does vanish at the endpoints and insert this into the second
variation (\ref{varent}). Integrating by parts, using the vanishing of $\delta m$ at the endpoints and
${\cal L}\chi_{_0} = 0$ gives
\begin{equation}
\delta^2 S= -{a k_{_B}\over \hbar G\,}\int_{r_1}^{r_2}r\, dr\, h^{-\frac{1}{2}}
\left(2\,{d m\over dr}\right)^{-{3\over 2}}
\chi_{_0}^2\,\left({d\psi\over dr}\right)^2 < 0\,.
\label{secvar}
\end{equation}
Thus the entropy of the solution is maximized with respect to radial
variations that vanish at the endpoints, {\it i.e.} those that do not
vary the positions of the interfaces. Perturbations of the
fluid in region II which are not radially symmetric decrease the entropy even
further than (\ref{secvar}), which demonstrates that the solution is stable to all small
perturbations keeping the endpoints fixed.

Allowing for endpoint variations as well requires the inclusion of the vacuum stress,
$\langle T_a^{\ b}\rangle$ in the vicinity of the interfaces, which fixed $\epsilon$ by the
estimate (\ref{epsest}). It is clear that the vacuum $\langle T_a^{\ b}\rangle$ must be included
in a more complete model for another reason. The general stress-energy in a spherically
symmetric, static spacetime has three components, namely $\rho, p$ and $p_{\perp}$.
We have set $p_{\perp} = p$ and restricted ourselves to only two isotropic eqs. of state
$p=-\rho$ and $p=+\rho$ only for simplicity, to illustrate the general features of a
non-singular solution to the gravitational collapse problem in a concrete example.
For any static solution, we must expect also the stress-energy tensor of vacuum polarization
in the Boulware vacuum to contribute. This stress-energy satisfies $p = \rho/3 <0$ near the
horizon \cite{CFul}. The addition of such a negative pressure eq. of state in the thin outer
edge obviates the need for the positive pressure discontinuity from negative pressure inside
to positive pressure outside.
Hence a completely smooth matching of $h$ and $df/dr$ is possible and the surface tensions
(\ref{surfa})-(\ref{surfb}) can be made to vanish identically. A full analysis of dynamical stability without
restriction on the interface boundaries will be possible in the framework of a more detailed model
which leads to these vacuum stresses in the boundary layer. Such an investigation can be carried
out without reference to thermodynamics or entropy and would apply then even in the case of a
configuration at absolute zero.

{\bf Conclusions.} A compact, non-singular solution of Einstein's eqs.
has been presented here as a possible stable alternative to black holes for the
endpoint of gravitational collapse. Realizing this alternative requires that a
quantum gravitational vacuum phase transition intervene before the classical event
horizon can form. Since the entropy of these objects is of the same order
as that of a typical stellar progenitor, even for $M > 100 M_{\odot}$, there is no
entropy paradox and no significant entropy shedding needed to produce a cold gravitational
vacuum or `grava(c)star' remnant.

Since the exterior spacetime is Schwarzschild until distances of order of the diameter of an
atomic nucleus from $r=r_{_S}$, a gravastar cannot be distinguished from a black
hole by present observations of X-ray bursts \cite{Abr}. However, the shell
with its maximally stiff eq. of state $p=\rho$, where the speed of sound is equal
to the speed of light, could be expected to produce explosive outgoing shock fronts in the
process of formation. Active dynamics of the shell may produce other effects that
would distinguish gravastars from black holes observationally, possibly providing
a more efficient particle accelerator and central engine for energetic astrophysical sources.
The spectrum of gravitational radiation from a gravastar should bear the imprint of
its fundamental frequencies of vibration, and hence also be quite different from a classical
black hole.

Quantitative predictions of such astrophysical signatures will require an
investigation of several of the assumptions, and extension of the simple model
presented in this paper in several directions. Although the eq. of state $p=\rho$ is strongly
suggested both by the limit of causality characteristic of a relativistic phase transition,
and by the correspondence of the fluid entropy with $S_{_{BH}}$ when the inner GBEC region is
shrunk to zero, this eq. of state has been assumed here, not derived from first principles.
Knowledge of the effective excitations in the shell is necessary to
determine the chemical potential $\mu$, and whether the entropy estimate (\ref{entest}) is
accurate, or more properly to be regarded as an upper bound on the entropy of a GBEC star.
The neglect of the $p= \rho/3 <0$ vacuum polarization in our model leads to some
freedom in matching at the two interfaces and the surface tensions (\ref{surfa})-(\ref{surfb}),
which may be different in detail or not present at all in a more complete treatment. A full analysis
of the dynamical stability of the object, including the motion of the interfaces or the boundary
layer(s) which replace them requires at least a consistent mean field description of quantum effects
in this transition region. Although general theoretical considerations indicate that
non-local quantum effects may be present in the vicinity of classical event horizons, a detailed
discussion of how these effects can alter the classical picture of gravitational collapse to a black
hole has not been attempted in this paper. Lastly, distinguishing the signatures of gravastars from
classical black holes in realistic astrophysical environments, such as in the presence of nearby
masses or accretion disks will depend on the details of the dynamical surface modes, as well as
the extension of the spherically symmetric static model presented here to include rotation and
magnetic fields.

One may regard the model presented in this paper as a proof of principle, the simplest
example of a physical alternative to the formation of a classical black hole, consistent
with quantum principles, which is free of any interior singularity or information paradox.
Additional theoretical and observational effort will be required to establish the cold,
dark, compact objects proposed in this paper as the stable final states of gravitational collapse.

Finally let us note that the interior de Sitter region with $p = -\rho$ may be interpreted
also as a cosmological spacetime, with the horizon of the expanding universe replaced by
a quantum phase interface. The possibility that the value of the vacuum energy density
in the effective low energy theory can depend dynamically on the state of a gravitational
condensate may provide a new paradigm for cosmological dark energy in the universe.
The proposal that other parameters in the standard model of particle physics may depend
on the vacuum energy density within a gravastar has been discussed by Bjorken \cite{BJ}.

Research of P. O. M. supported in part by NSF grant 0140377.


\begin{thebibliography}{99}

\centerline{\bf{References}}
\vspace{.5cm}
\bibitem{Bou75} Boulware, D. G. (1975) {\it Phys. Rev.} {\bf D11}, 1404-1423.
\bibitem{Bou76} Boulware, D. G. (1976) {\it Phys. Rev.} {\bf D13}, 2169-2187.
\bibitem{CFul} Christensen, S. M. \& Fulling, S. A. (1977) {\it Phys. Rev.} {\bf D15}, 2088-2104.
\bibitem{Bek72} Bekenstein, J. D. (1972) {\it Nuovo Cimento Lett.} {\bf 4}, 737-740.
\bibitem{Bek73} Bekenstein, J. D. (1973) {\it Phys. Rev.} {\bf D7}, 2333-2346.
\bibitem{Bek74} Bekenstein, J. D. (1974) {\it Phys. Rev.} {\bf D9}, 3292-3300.
\bibitem{Haw74} Hawking, S. W. (1974) {\it Nature} {\bf 248}, 30-31.
\bibitem{Haw75} Hawking, S. W. (1975) {\it Comm. Math. Phys.} {\bf 43}, 199-220.
\bibitem{Chr} Christodolou, D. (1970) {\it Phys. Rev. Lett.} {\bf 25}, 1596-1597.
\bibitem{ChrRuf} Christodolou D. \& Ruffini, R. (1971) {\it Phys. Rev.} {\bf D4}, 3552-3555.
\bibitem{HawEll} Hawking, S. W. \& Ellis, G. F. R. (1973) {\it The Large Scale Structure of
Space-Time} \hfill\break (Cambridge Univ. Press, Cambridge).
\bibitem{Haw76} Hawking, S. W. (1976) {\it Phys. Rev.} {\bf D13}, 191-197.
\bibitem{tH85} `t Hooft, G. (1985) {\it Nucl. Phys.} {\bf B256}, 727-745.
\bibitem{BKLS} Bombelli, L., Koul, R. K., Lee, J \& Sorkin, R. (1986) {\it Phys. Rev.} {\bf D34},
373-383.
\bibitem{Haw82} Hawking, S. W. (1982) {\it Comm. Math. Phys.} {\bf 87}, 395-415.
\bibitem{tH99} `t Hooft, G. (1999) {\it Class. Quan. Grav.} {\bf 16}, 3263-3279.
\bibitem{Maz} Mazur, P. O. (1996) {\it Acta Phys. Pol.} {\bf B27}, 1849-1858; See also
Gorski, A. Z. \& Mazur, P. O., e-print arXive: hep-th/9704179.
\bibitem{HarHaw} Hartle J. B. \& Hawking, S. W. (1976) {\it Phys. Rev.} {\bf D13}, 2188-2203.
\bibitem{ZP} Zurek, W. H. \& Page, D. N. (1984) {\it Phys. Rev.} {\bf D29}, 628-631.
\bibitem{tH98} 't Hooft, G., (1998) {\it Nucl. Phys. Proc. Suppl.} {\bf 68}, 174-184.
\bibitem{CHLS01} Chapline, G., Hohlfield, E., Laughlin, R. B. \& Santiago, D. I. (2001)
{\it Phil. Mag.} {\bf B81}, 235-254.
\bibitem{Lau} Laughlin, R. B. (2003) {\it Int. Jour. Mod. Phys.} {\bf A18}, 831-853.
\bibitem{CHLS03} Chapline, G., Hohlfield, E., Laughlin, R. B. \& Santiago, D. I. (2003)
{\it Int. Jour. Mod. Phys.} {\bf A18}, 3587-3590.
\bibitem{Gli} Gliner, {\'E}. B. (1965) {\it Hz. Eksp. Teor. Fiz.} {\bf 49}, 542-548
[(1966) {\it Sov. Phys. JETP} {\bf 22}, 378-382].
\bibitem{Sak} Sakharov, A. D. (1965) {\it Hz. Eksp. Teor. Fiz.} {\bf 49}, 345-358
[(1966) {\it Sov. Phys. JETP} {\bf 22}, 241-249].
\bibitem{Chap} Chapline, G. (1992) in {\it Foundations of Quantum Mechanics}, eds. Black, T. D.,
Nieto, M. M., Pilloff, H. S., Scully, M. O. \& Sinclair, R. M., pp. 255-260
(World Scientific, Singapore).
\bibitem{MMPRD} Mazur P. O. \& Mottola, E. (2001) {\it Phys. Rev.} {\bf D64}, 104022,
and references therein.
\bibitem{Dym} Dymnikova, I. (2003) {\it Int. Jour. Mod. Phys.}, {\bf D12}, 1015-1034.
\bibitem{Isr} Israel, W. (1966) {\it Nuovo Cimento} {\bf B44}, 1-14; {\bf B48}, 463.
\bibitem{ZelNov} Zel'dovich, Ya. B. \& Novikov, I. D. (1971) {\it Relativistic Astrophysics}, Vol. 1,
(University of Chicago Press, Chicago) [(1996) {\it Stars and Relativity} (Dover, New York)].
\bibitem{Coc} Cocke, W. J. (1965) Ann. Inst. H. Poincar{\'e} {\bf A2}, 283-306.
\bibitem{Abr} Abramowicz, M. A., Kluzniak, W. \& Lasota, J.-P. (2002) {\it Astron. Astroph.},
{\bf 396}, L31-L34.
\bibitem{BJ} Bjorken, J. D. (2003) {\it Phys. Rev.} {\bf D67}, 043508.
\end{thebibliography}
\end{document}